\documentclass[prb,aps,twocolumn,amsmath,footinbib,superscriptaddress,10pt]{revtex4-2}
\usepackage[utf8]{inputenc}
\usepackage{amssymb,amsmath}
\usepackage{graphicx}
\usepackage{natbib}
\usepackage{multirow}
\usepackage{physics}
\usepackage{natbib}
\usepackage{cancel}

\usepackage{xcolor}

\usepackage[caption=false]{subfig}
\newcommand{\phantomsubfloat}[1]{
    {% apply caption setup only temporarily
        \captionsetup[subfigure]{labelformat=empty}
        \subfloat[][]{#1}
    }%
}

\usepackage[final]{hyperref} % adds hyper links inside the generated pdf file
\hypersetup{
	colorlinks=true,       % false: boxed links; true: colored links
	linkcolor=red,        % color of internal links
	citecolor=red,        % color of links to bibliography
	filecolor=magenta,     % color of file links   
	urlcolor=blue         
}

\usepackage{color} % for colored text

\definecolor{darkgreen}{rgb}{0,0.7,0}  
\definecolor{orange}{RGB}{255,127,0}

\begin{document}

\title{Anomalous charge transport in the sine--Gordon model}

\author{Frederik M{\o}ller}\email{frederik.moller@tuwien.ac.at}
\affiliation{
Vienna Center for Quantum Science and Technology (VCQ), Atominstitut, TU Wien, Vienna, Austria}
\author{Botond C. Nagy}\email{botond.nagy@edu.bme.hu}
\affiliation{Department of Theoretical Physics, Institute of Physics, Budapest University of Technology and Economics, H-1111 Budapest, M{\H u}egyetem rkp.~3.}
\affiliation{
BME-MTA Momentum Statistical Field Theory Research Group, Institute of Physics, Budapest University of Technology and Economics, H-1111 Budapest, M{\H u}egyetem rkp.~3.}
\author{M\'arton Kormos}\email{kormos.marton@ttk.bme.hu}
\affiliation{Department of Theoretical Physics, Institute of Physics, Budapest University of Technology and Economics, H-1111 Budapest, M{\H u}egyetem rkp.~3.}
\affiliation{MTA-BME Quantum Correlations Group (ELKH), Institute of Physics, Budapest University of Technology and Economics, H-1111 Budapest, M{\H u}egyetem rkp.~3.}
\affiliation{
BME-MTA Momentum Statistical Field Theory Research Group, Institute of Physics, Budapest University of Technology and Economics, H-1111 Budapest, M{\H u}egyetem rkp.~3.}
\author{G\'abor Tak\'acs}\email{takacs.gabor@ttk.bme.hu}
\affiliation{Department of Theoretical Physics, Institute of Physics, Budapest University of Technology and Economics, H-1111 Budapest, M{\H u}egyetem rkp.~3.}
\affiliation{MTA-BME Quantum Correlations Group (ELKH), Institute of Physics, Budapest University of Technology and Economics, H-1111 Budapest, M{\H u}egyetem rkp.~3.}
\affiliation{
BME-MTA Momentum Statistical Field Theory Research Group, Institute of Physics, Budapest University of Technology and Economics, H-1111 Budapest, M{\H u}egyetem rkp.~3.}

\date{November 16, 2024}
%Version date

\begin{abstract} 

We conduct a comprehensive study of anomalous charge transport in the quantum sine--Gordon model.
Employing the framework of Generalized Hydrodynamics, we compute Drude weights and Onsager matrices across a wide range of coupling strengths to quantify ballistic and diffusive transport, respectively.
We find that charge transport is predominantly diffusive at accessible timescales, indicated by the corresponding Onsager matrix significantly exceeding the Drude weight -- contrary to most integrable models where transport is primarily ballistic.
Reducing the Onsager matrix to a few key two-particle scattering processes enables us to efficiently examine transport in both low- and high-temperature limits.
The charge transport is dictated by non-diagonal scattering of the internal charge degree of freedom: 
At particular values of the coupling strength with diagonal, diffusive effects amount to merely subleading corrections.
However, at couplings approaching these points, the charge Onsager matrix and corresponding diffusive time-scale diverge.
Our findings relate to similar transport anomalies in XXZ spin chains, offering insights through their shared Bethe Ansatz structures.
\end{abstract} 

\maketitle 

\section{Introduction}

With recent advancements in the realization and manipulation of quantum many-body systems -- particularly in the domain of ultracold atoms~\cite{2008RvMP...80..885B} -- there has been a growing demand for frameworks to describe their transport and far-from-equilibrium dynamics.
Integrable models are particularly notable in this context, as they are some of the few strongly interacting quantum systems for which exact results are available, providing a unique perspective for examining transport phenomena \cite{2021RvMP...93b5003B}. These systems have an extensive number of conserved quantities and stable quasi-particle excitations, which can lead to anomalous transport properties \cite{2016JSMTE..06.4010V,2021RvMP...93b5003B}. Although quasi-particle propagation is mainly ballistic, reflected in their finite Drude weights \cite{1995PhRvL..74..972C,2017PhRvL.119b0602I} detected by the Mazur inequality \cite{1969Phy....43..533M,1971Phy....51..277S}, integrable systems can also feature diffusive \cite{Medenjak2017,2018PhRvB..98v0303G} and even superdiffusive transport \cite{2019PhRvL.122u0602L,Gopalakrishnan2019,2021JSMTE2021h4001B,2021PhRvX..11c1023I}.

A significant theoretical development in this area is Generalized Hydrodynamics (GHD)~\cite{2016PhRvX...6d1065C, 2016PhRvL.117t7201B}, a recent hydrodynamic approach specifically designed for integrable systems, which facilitates a detailed analysis of transport and reveals connections between microscopic processes and emergent macroscopic behavior (see reviews~\cite{Cubero_2021, 2021JSMTE2021i4001B, 2022JSMTE2022a4002D, Bulchandani2021, 2023arXiv231103438D}).
Although integrability imposes specific, idealized scattering conditions, several experimentally realized systems are approximately integrable and can therefore be described by GHD~\cite{2019PhRvL.122i0601S, doi:10.1126/science.abf0147, 2021PhRvL.126i0602M, 2020Sci...367.1461W, 2022PhRvX..12d1032C, PhysRevLett.133.113402, doi:10.1126/science.adk8978, arXiv:2406.17569}. 

In this work, we perform an in-depth study of transport in the sine--Gordon (sG) quantum field theory, a model with numerous applications as the low-energy description of a large spectrum of gapped one-dimensional systems \cite{Giamarchi:743140}, ranging from quasi-1D antiferromagnets and carbon nanotubes through organic conductors \cite{Controzzi2001,2005ffsc.book..684E}, trapped ultra-cold atoms \cite{2007PhRvB..75q4511G,2010PhRvL.105s0403C,2010Natur.466..597H,2017Natur.545..323S,PRXQuantum.4.030308}, to quantum circuits \cite{2021NuPhB.96815445R} and coupled spin chains \cite{Wybo2022}. 
It is an integrable model with a known exact $S$-matrix \cite{1977CMaPh..55..183Z,ZAMOLODCHIKOV1979253}, thus facilitating treatment via the Bethe Ansatz and GHD.
Nevertheless, the current understanding of the general transport properties of the quantum sine--Gordon model remains limited, as its hydrodynamic description for generic couplings has only been obtained recently \cite{2023PhRvB.108x1105N,2024ScPP...16..145N}.

Previous works on transport in the sine-Gordon model studied the optical conductivity using spectral expansion techniques \cite{2001PhRvL..86..680C} based on the exact solution of the form factor bootstrap \cite{1992ASMP...14....1S}. More recently, it was shown that the Drude weight characterizing the ballistic charge transport displays a fractal structure~\cite{2023PhRvB.108x1105N}, similar to the spin Drude weight in the gapless XXZ spin chain~\cite{2013PhRvL.111e7203P, 2017PhRvL.119b0602I, 2019PhRvL.122o0605L, 2020PhRvB.101v4415A, 2022JPhA...55X4005I}. The origin of this structure is the presence of reflective scattering, which also results in other anomalous transport properties such as separation of the charge and energy transport \cite{2024PhRvB.109p1112M}.

The fractal structure of the Drude weight indicates that the low-frequency charge transport is anomalous, motivating us to examine the roles of diffusive processes. Within the GHD framework, one-particle processes dictate the ballistic dynamics at the lowest order hydrodynamic expansion, while two-particle processes result in diffusive corrections~\cite{2018PhRvL.121p0603D, 2019ScPP....6...49D}.
We demonstrate that following the non-diagonal scattering of the internal charge degree of freedom, diffusive spreading of charge is strongly enhanced and exhibits anomalous scaling. We also connect our findings to studies conducted in the semi-classical limit~\cite{2005PhRvL..95r7201D, 2019PhRvB.100c5108B} and of the closely related XXZ spin chains~\cite{2018PhRvB..97h1111C, 2020PhRvB.101v4415A}, thereby providing a comprehensive picture of transport.

The paper is structured as follows.
In Sec.~\ref{sec:SG_theory}, we introduce the quantum sine--Gordon model, focusing on the scattering properties of its excitations and how they manifest in the thermodynamics and hydrodynamics of the model.
In Sec.~\ref{sec:transport_coeffs}, we establish the transport coefficients studied throughout the paper and demonstrate their anomalous scaling with the coupling strength of the model.
Next, in Sec.~\ref{sec:scattering_contributions}, we break down the individual scattering contributions to diffusive transport and illustrate how they scale with temperature.
Building on this, Secs.~\ref{sec:highT} and~\ref{sec:lowT} treat the high- and low-temperature limits of charge transport; the former focuses on the divergence of the Onsager matrix when adding magnonic excitations to the spectrum, while the latter explores the regime approaching the classical limit of vanishing coupling strength.
In Sec.~\ref{sec:finite_scales}, we consider the crossover between diffusive and ballistic transport at finite time scales in the bipartition protocol, both in the hydrodynamic and microscopic picture.
Finally, Sec.~\ref{sec:conclusions} contains our conclusions and outlook for future studies.
Several appendices containing more technical aspects complement the main text.

\section{sine--Gordon hydrodynamics} \label{sec:SG_theory}

The sine--Gordon model is a relativistic field theory with Hamiltonian
\begin{equation}
    H=\int \mathrm{d} x\left[\frac{1}{2}\left(\partial_t \phi\right)^2+\frac{1}{2}\left(\partial_x \phi\right)^2-\lambda \cos (\beta \phi)\right] \text {, }
\end{equation}
where $\phi(x)$ is a real scalar field, $\beta$ is a dimensionless coupling strength, and the dimensionful parameter $\lambda$ sets the mass scale. 

The fundamental excitations are topologically charged \textit{kinks/antikinks} of mass $m_S$, which interpolate between the degenerate vacua of the periodic cosine potential.
To describe the spectrum and the scattering of excitations, it is useful to introduce the renormalized coupling constant
\begin{equation}
    \xi = \frac{\beta^2}{8 \pi - \beta^2}\,.
\end{equation}
In the attractive regime $0 < \xi < 1$ kink-antikink pairs can form neutral bound states dubbed \textit{breathers} $B_k$, with masses 
\begin{equation}
    m_{B_k} = 2 m_S \sin \left( \frac{k \pi \xi}{2} \right)\,,
\end{equation} 
where $k$ runs from $1$ to $N_B=\lfloor 1 / \xi\rfloor$.
It is convenient to use units setting $\hbar=1$ and the speed of light (the speed of sound in condensed matter context) $c=1$, as well as to set the Boltzmann constant $k_B=1$.
As a result, energies and temperatures are measured in units of $m_S$, while distances and times are measured in units of $1/m_S$. 
The energy $E$ and momentum $p$ of excitations $a$ with mass $m_a$ can be parameterized by the rapidity variable $\theta$ as $E=m_a\cosh\theta$ and $p=m_a\sinh\theta$.

\subsection{Scattering amplitudes and Bethe Ansatz}\label{subsec:TBA}

To understand the transport properties of the sine--Gordon model, it is necessary to first understand its spectrum of excitations and their scattering.
While most scattering processes are purely transmissive, the kink-antikink scattering can be both transmissive and reflective with respective amplitudes given by 
\begin{align}
    S_T(\Delta\theta)&=\frac{\sinh \left( \Delta\theta /\xi \right)}{\sinh \left(( i \pi-\Delta\theta)/\xi\right)} S_0(\Delta\theta,\xi)\, , \\
    S_R(\Delta\theta)&=\frac{i \sin \left(\pi/\xi\right)}{\sinh \left((i \pi-\Delta\theta )/\xi\right)}  S_0(\Delta\theta,\xi)\, , \label{eq:reflective_amplitude}
\end{align}
where $\Delta\theta$ is the rapidity difference between the excitations and $S_0$ is a phase factor. 
For integer values of $1/\xi$, the reflective scattering amplitude vanishes; for these particular couplings, referred to as \textit{reflectionless points}, the physical particles (kinks, antikinks, and breathers) constitute the entire excitation spectrum.

The presence of reflective scattering significantly alters the transport of topological charge.
To fully appreciate this, it is instructive to first consider the semi-classical approach of Ref.~\cite{2016PhRvE..93f2101K}, which assumes the momenta of all quasi-particles to be negligibly small and that a Dirac-delta can approximate the scattering potential.
Hence, the scattering matrix becomes independent of the incoming momenta and fully reflective in the space of internal quantum numbers, i.e., the topological charge.
Figure~\ref{fig:propagation_illustration_a} illustrates the resulting dynamics of a single antikink in a background of kinks. 
Although the particles propagate ballistically, the reflective scattering inhibits ballistic transport of charge, which spreads diffusively instead.
Conserved quantities unrelated to the internal quantum numbers, such as momentum and energy, are unaffected by the reflections and thus remain transported ballistically.

Beyond the semi-classical approach, the sine--Gordon model can be solved exactly via the Bethe Ansatz, which parameterizes the solution of integrable systems in terms of quasi-particle excitations~\cite{1969JMP....10.1115Y, takahashi_1999}.
Due to the non-diagonal scattering, the Bethe Ansatz equations are nested, requiring an auxiliary Bethe Ansatz system for the internal degrees of freedom.
Notably, the auxiliary Bethe equations are the same as the Bethe equations of the XXZ spin chain in its gapless (easy-plane) phase.
Ultimately, the thermodynamic description can be formulated in terms of quasi-particle excitations consisting of the breathers $B_k$, a single solitonic excitation $S$ accounting for the energy and momentum of the kinks, and also partly for the charge, and additional massless auxiliary excitations, dubbed \textit{magnons}, which account for the internal degeneracies related to the charge degrees of freedom.
Physically, a magnon corresponds to a ``topological-charge flip" relative to an all-solitonic reference state, similar to the scenario depicted in Fig.~\ref{fig:propagation_illustration_a}.

In the thermodynamic limit, the roots of the magnonic Bethe equations arrange themselves into ``strings" (a set of complex roots with the same real part), with each string corresponding to a bound state of elementary magnons. The different strings, including the elementary magnons, are called separate magnon species.
Due to the mapping to the XXZ chain, these solutions correspond to the gapless XXZ spin chain strings. Crucially, the soliton, the breathers, and the magnons all have \emph{diagonal} scattering.

The magnon species can be classified by writing the coupling $\xi$ as a continued fraction
\begin{equation}
    \xi = \frac{1}{\displaystyle N_B+\frac{1}{\displaystyle\nu_1+\frac{1}{\displaystyle\nu_2 + \ldots}}}\,,
\end{equation}
with $N_B$ breathers and $\nu_k$ different magnon species at level $k$.
Following the thermodynamic Bethe Ansatz, the system can be described in terms of the total densities of states $\rho_a^{\text{tot}} (\theta)$ and densities of occupied states $\rho_a (\theta)$ of quasi-particle excitation with species $a$.

\subsection{Soliton gas picture for quasi-particle dynamics}\label{subsec:soliton_gas}

\begin{figure}
    \centering
    \includegraphics[width=1.0\columnwidth]{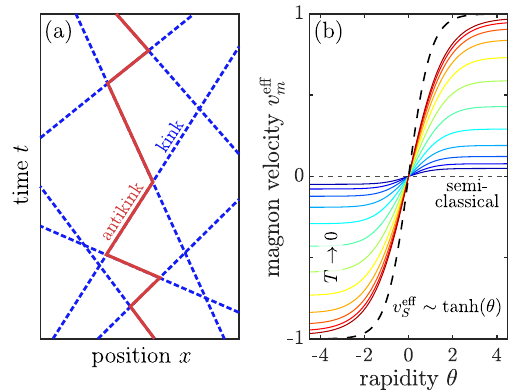}
    \phantomsubfloat{\label{fig:propagation_illustration_a}}
    \phantomsubfloat{\label{fig:propagation_illustration_b}}
    \vspace{-2\baselineskip}% Remove extra line inserted by subfloat
    
    \caption{Illustration of excitation dynamics. (a) Semi-classical particle trajectories of a single antikink in a background of kinks. Due to their reflective scattering, the charge carried by the antikink spreads diffusively. (b) Effective magnon velocities for different temperatures $T$. For $T \to 0$, GHD approaches the semi-classical approximation as the ballistic magnon velocity vanishes.}
    \label{fig:propagation_illustration}
\end{figure}

At the lowest order of hydrodynamic expansion, the Euler scale, Generalized Hydrodynamics describes the evolution of integrable systems via the collisionless Boltzmann equations~\cite{2016PhRvX...6d1065C, 2016PhRvL.117t7201B}  
\begin{equation}
    \partial_t \rho_a(\theta, x, t) + \partial_x \left(v_{a}^{\mathrm{eff}}(\theta, x, t) \: \rho_a(\theta, x, t) \right) = 0 \, ,
    \label{eq:Euler_GHD}
\end{equation}
where the effective velocity $v_{a}^{\mathrm{eff}}(\theta)$ depends on local interactions with other particles~\cite{2014PhRvL.113r7203B,2023arXiv230709307D} (see Appendix for explicit expression).
A particularly useful physical interpretation of Eq.~\eqref{eq:Euler_GHD} derives from its mapping to a classical soliton gas~\cite{2018PhRvB..97d5407B, PhysRevLett.120.045301}: 
Each quasi-particle of species $a$ with rapidity $\theta$ propagates ballistically along classical trajectories with a time-independent velocity.
Although 
%akin to the semi-classical approach, 
all collisions are transmissive in the magnonic picture,
elastic collisions with other particles result in the particles acquiring a Wigner time delay dependent on the scattering phase factor; 
the accumulated contributions of all delays yields the effective velocity $v_{a}^{\mathrm{eff}}(\theta)$. 
As a result, while the Boltzmann equation includes no collision integral term in the usual sense, the effective velocity does account for the effect of quasi-particle collision.

However, despite the diagonal quasi-particle scattering, signs of the underlying partially reflective scattering of the physical particles can be seen in the effective velocities of magnons, particularly at lower temperatures.
This is illustrated in Fig.~\ref{fig:propagation_illustration_b}, depicting the magnon and soliton velocities at different temperatures for a neutral system in thermal equilibrium.
In the low-temperature limit, i.e.\ at temperatures far below the gap $T \ll m_{B_1}$, the magnons become non-dispersive, as their effective velocity for all rapidity states tends toward the solitonic fluid velocity $v_S = J_s /n_S$, where $J_S = \int \mathrm{d}\theta \, \rho_S(\theta) v_{S}^{\mathrm{eff}}(\theta)$ and $n_S = \int \mathrm{d}\theta \, \rho_S(\theta) $ is the current density and linear density of solitons, respectively.
Thus, given a finite $J_S$, the magnons are dragged along with the solitons, while for $J_S = 0$ the ballistic magnon velocity vanishes, akin to the semi-classical scenario depicted in Fig.~\ref{fig:propagation_illustration_a}.
Indeed, as shown analytically in Ref.~\cite{2019PhRvB.100c5108B}, predictions of the semi-classical approach are recovered in the low-temperature limit of GHD.
Away from this limit, however, charge transport is always ballistic.
Nevertheless, the magnons continue to experience a significant drag by the solitons, resulting in transport phenomena such as dynamical charge-energy separation and distinctively shaped “arrowhead” light cones~\cite{2024PhRvB.109p1112M}.
Finally, in the high-temperature limit, the system decouples into independent left and right moving modes.
Hence, the effective velocities of the magnons approach that of the solitons.
We study this limit in further detail below in Section~\ref{sec:highT}.

\subsection{Diffusive corrections}

The next order of hydrodynamic expansion accounts for dissipative effects, which in GHD manifests as a diffusive broadening of the ballistic quasi-particle trajectories~\cite{2018PhRvL.121p0603D, 2019ScPP....6...49D, 2018PhRvB..98v0303G}.
Again, the soliton gas picture offers an appealing physical interpretation: 
Gaussian fluctuations of the quasi-particle densities $\rho_a$ result in fluctuations of the number of collisions experienced by a particle propagation through a region of the system.
Hence, the accumulated Wigner delay fluctuates, resulting in the trajectory of each particle broadening as it follows a biased random walk.
In turn, the ballistic transport of conserved quantities, which are carried by the quasi-particle excitations, pick up diffusive corrections as well.

Unlike diffusion in non-integrable models, diffusion in GHD arises mostly as a subleading correction to the ballistic Euler-scale hydrodynamics.
Integrable diffusion is nevertheless important, resulting in effects such as an effective viscosity at low temperature~\cite{PhysRevLett.132.243402} and thermalization in the presence of an external potential~\cite{PhysRevLett.125.240604, PhysRevResearch.6.023083}.
The role of higher-order hydrodynamic terms is yet to be fully understood~\cite{PhysRevB.96.220302, 2023JPhA...56x5001D, PhysRevResearch.6.013328}.

The variance of each quasi-particle trajectory, and thus the diffusive spreading of different quantities, depends on the scattering properties of the individual particle species.
A detailed analysis of two-body scattering processes in the sine--Gordon model is conducted in Sec.~\ref{sec:scattering_contributions}.
In relativistically invariant field theories, energy spreading is non-diffusive~\cite{2019ScPP....6...49D}.
Furthermore, no cross effects exist between topological charge, momentum, and energy at zero chemical potential~\cite{2024ScPP...16..145N}.
Therefore, we will only consider diagonal topological charge and momentum transport values.

\section{Coefficients of transport} \label{sec:transport_coeffs}

\begin{figure*}
    \centering
    \includegraphics[width=1.0\textwidth]{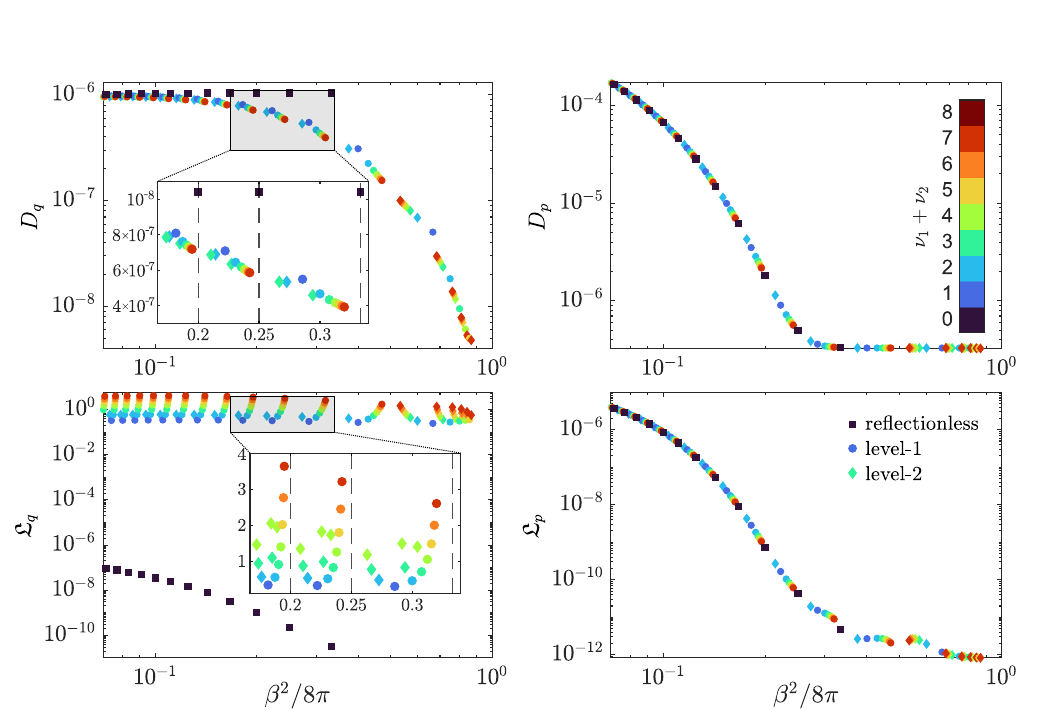}
    \caption{Charge and momentum transport coefficients for $T = 0.1$. Each point is computed for a separate coupling and has a unique quasi-particle spectrum; the shape of the point indicates the number of magnon levels, while its color denotes the total number of magnon species. 
    In (a) and (b), the insets display a zoomed-in region plotted with linear axes.}
    \label{fig:transport_coefficients}
\end{figure*}

Two transport coefficients, instrumental in characterizing the transport properties of many-body systems, are the Drude weight $D_i$ and Onsager matrix $\mathfrak{L}_i$.
The subscript $i$ refers to a conserved quantity whose linear density and associated current we denote by $n_i$ and $j_i$, respectively.
Conventionally, the real part of the corresponding conductivity $\sigma_{i}$ takes the form
\begin{equation}
    \mathrm{Re}\,\sigma_{i}(\omega)=\pi T^{-1} D_{i}\,\delta(\omega) + \sigma^\text{reg.}_{i}(\omega)\,,
\end{equation}
composed of a singular contribution quantified by the Drude weight $D_{i}$ and a remaining frequency-dependent part. 
The d.c.\ conductivity is obtained in the $\omega \to 0$ limit of the conductivity after the Drude weight is subtracted and is related to the Onsager matrix $\mathfrak{L}_{i}$ via
\begin{equation}
    \mathfrak{L}_{i}=(2 T)\lim_{\omega\to0}\sigma^\text{reg.}_{i}(\omega) \; .
\end{equation} 
Both transport coefficients are typically expressed in terms of the time-averaged current–current auto-correlation function $C_i (x, t) \equiv \langle j_i(x,t) j_i(0,0)\rangle^c$ as
\begin{align}
    D_{i}&={\lim_{\tau\to\infty}} \frac{1}{2\tau}\int_{-\tau}^{\tau} \mathrm{d}t \int \mathrm{d}x \,C_i (x, t)\,,\\
    \mathfrak{L}_{i} &= \lim_{\tau\to\infty} \int_{-\tau}^{\tau} \mathrm{d}t \left(\int \mathrm{d}x \,C_i (x, t)-D_{i} \right)\,.
\end{align}
Hence, a finite Drude weight $D_{i} > 0$ indicates a dissipationless current $j_i$ and thus ballistic transport.

In the context of hydrodynamic expansions, the Drude weight and the Onsager matrix provide complementary perspectives on transport, each associated with different orders of the expansion~\cite{De_Nardis_2022}. 
The Drude weight is tied to ballistic transport (the Euler-scale hydrodynamics), while the Onsager matrix incorporates dissipative effects, here in the form of diffusion.
Remarkably, the GHD picture yields exact expressions of the two transport coefficients in integrable systems through form factor expansion of quasi-particle excitations on top of the local steady state.
The resulting Drude weight is completely determined by single-particle processes and reads~\cite{Doyon2017b}
\begin{equation}
    D_i = \sum_a \int \text{d}\theta\ \chi_a(\theta) \left(h_{i;a}^{\text{dr}}(\theta) \: v_a^{\text{eff}}(\theta)\right)^2\, ,
    \label{eq:Drude}
\end{equation}
whereas only two-particle scattering processes contribute to the diffusive dynamics and thus the Onsager matrix~\cite{2019ScPP....6...49D}
\begin{equation}
\begin{aligned}
    \mathfrak{L}_{i} =&\sum_{a,b}\int\frac{\text{d}\theta_1\text{d}\theta_2}{2} \chi_a(\theta_1) \chi_b(\theta_2) \left| v_a^{\text{eff}}(\theta_1) - v_b^{\text{eff}}(\theta_2) \right| \times \\ 
    &\times \left( \frac{\Phi_{ab}^{\text{dr}}(\theta_1 -\theta_2)\eta_b}{2\pi} \right)^2 \left( \frac{h_{i;b}^{\text{dr}}(\theta_2)}{\rho_b^{\text{tot}}(\theta_2)} - \frac{h_{i;a}^{\text{dr}}(\theta_1)}{\rho_a^{\text{tot}}(\theta_1)} \right)^2 \,.
\label{eq:Onsager}
\end{aligned}
\end{equation}
Here, $\chi_a = \rho_a \left( 1 - \rho_a / \rho_{a}^{\mathrm{tot}} \right)$ is the quasi-particle susceptibility,  $\eta_a=\pm 1$ are appropriate sign factors, and $h_{i;a} (\theta)$ are the \textit{bare} values of conserved quantity $i$ carried by a particle species $a$ with rapidity $\theta$.
The superscript `dr' indicates that the quantity is \textit{dressed} modifying it through quasi-particle interactions (see Appendix).
Crucially, although the soliton and all magnons carry a bare topological charge, following dressing all charge is concentrated in the last two magnon species.

Employing Eqs.~\eqref{eq:Drude} and \eqref{eq:Onsager}, we calculate the transport coefficients of charge and momentum for thermal states at temperature $T=0.1$ for various couplings~\cite{2020ScPP....8...41M, 2023JCoPh.49312431M}.
The results are shown in Fig.~\ref{fig:transport_coefficients}.
Comparing the transport coefficients for topological charge and momentum, we observe very different behavior.
Whereas the momentum Drude weight $D_p$ (and the corresponding Onsager matrix $\mathfrak{L}_p$) is continuous as a function of the coupling, its charge counterpart exhibits a fractal structure, where $D_q$ can jump by $\mathcal{O}(1)$ for an infinitesimal change of $\beta$.
As shown in studies of the easy-plane XXZ spin chain~\cite{2020PhRvB.101v4415A}, which displays similar properties in its spin transport, these jumps in the zero-frequency spectral weight suggest a non-trivial behavior of the finite-frequency conductivity.
Indeed, the charge Onsager matrix plotted in Fig.~\ref{fig:transport_coefficients} exhibits several anomalous features.
First, while $\mathfrak{L}_q$ of the reflectionless points is of similar magnitude to the charge Drude weight, attractive points with just two level-1 magnons have $\mathfrak{L}_q \sim \mathcal{O}(1)$, i.e. several orders of magnitude higher.
This suggests that reflective scattering between excitations significantly alters the d.c. conductivity of charge.
Secondly, approaching one of the reflectionless points, i.e.\ adding additional magnons to the spectrum, leads to a divergence of $\mathfrak{L}_q$.
Such divergence of the Onsager matrix is associated with the breakdown of the diffusive expansion, in which case the model is expected to display super-diffusion~\cite{Bulchandani2021}.
This is similar to the behavior observed in Ref.~\cite{2020PhRvB.101v4415A} for the XXZ spin chain, where the anomalous response was interpreted as a consequence of quasi-particles undergoing Lévy flights; we shall return to this point in Section~\ref{sec:highT} where a detailed analysis of the divergence is performed.

\section{Scattering contributions to diffusive transport} \label{sec:scattering_contributions}

\begin{figure*}
    \centering
    \includegraphics[width=1.0\textwidth]{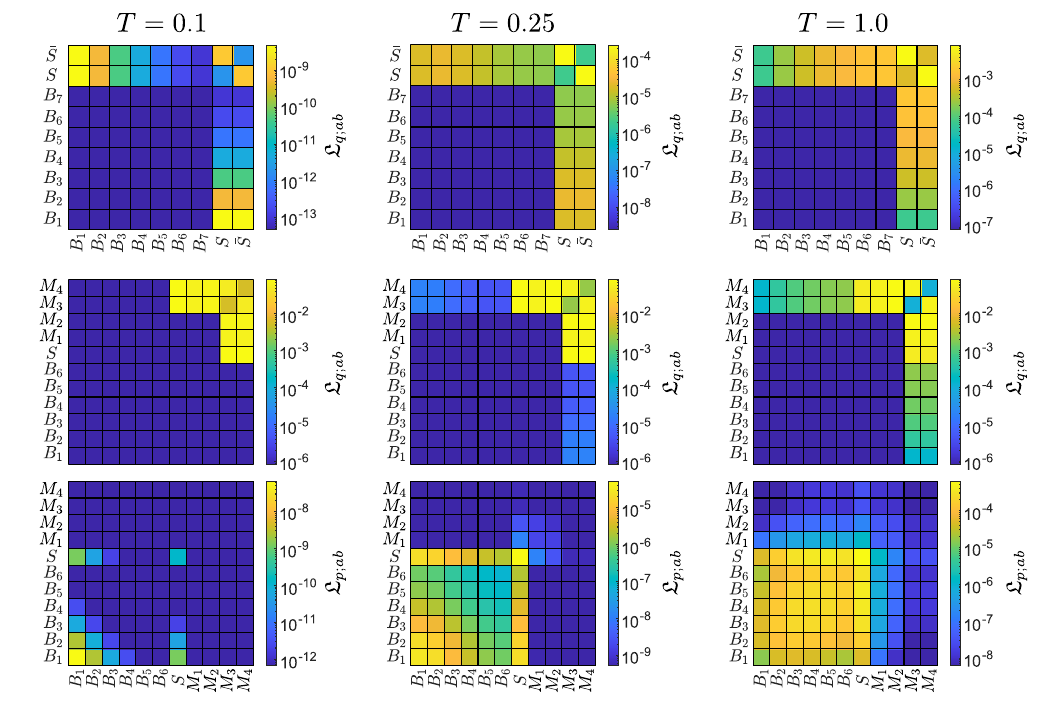}
    \caption{Contribution of each two-particle scattering to the Onsager matrix. The top row depicts the decomposed charge Onsager matrix for the reflectionless point $\xi = 1/8$, computed for three different temperatures. The bottom two rows are the charge and momentum matrices for the attractive point $\xi = 1/(6 + 1/4)$.}
    \label{fig:Onsager_particlewise}
\end{figure*}

According to form factor expansions, only two-particle scattering processes contribute to the diffusive GHD dynamics~\cite{2018PhRvL.121p0603D, 2019ScPP....6...49D}.
Hence, to gain physical insight into the anomalous transport behavior of the sine--Gordon model, we may study the individual two-body contributions to the Onsager matrix $\mathfrak{L}_{i} \equiv \frac{1}{2}\sum_{a,b} \mathfrak{L}_{i;a b} $. 
For simplicity, we restrict our analysis to a single level of magnons. 

Fig.~\ref{fig:Onsager_particlewise} shows results for different temperatures at two coupling points: a reflectionless point with seven breather species and an attractive point with six breathers and four magnon species.
In both cases, only specific particle species carry dressed topological charge: kinks/antikinks at the reflectionless couplings and the last two magnon species at generic couplings.
Consequently, only scattering events involving these charged particles contribute to the charge Onsager matrix.
Despite this commonality, the two regimes differ significantly.

At the reflectionless point, all quasi-particles are massive, and scattering shifts associated with heavier particles are generally larger.
However, at lower temperatures, the population of heavier particles is suppressed, whereby for $T = 0.1 \sim m_{B_1}/4$, scattering of the lightest breathers with kinks/antikinks is the dominant contributor.
As temperature increases, the heavier particles contribute more despite being less abundant.
Additionally, since kinks and antikinks carry topological charge, their mutual scattering plays a larger role than interactions with neutral particles.

In contrast, at the generic points, scattering of the dressed charge carriers (two last magnon species) and particles with a bare charge (magnons and the soliton) all make significant contributions to the charge Onsager matrix.
These scattering events contribute much more than any interactions involving neutral particles.
Hence, the generic points feature much higher diffusive charge transport than the reflectionless points.
Furthermore, unlike the scattering with massive neutral particles, the magnon-magnon and magnon-soliton scattering exhibit a much weaker temperature dependence.
Hence, the charge Onsager matrix only varies modestly with temperature, unlike other transport coefficients, such as the charge Drude weight, which exhibits a strong temperature dependence.

For the momentum Onsager matrix, the reflectionless and generic points show similar behavior (only the generic case is shown in Fig.~\ref{fig:Onsager_particlewise}).
Since magnons are massless and only have small dressed momenta, the main contributions come from scattering between massive particles.
At low temperatures, only the lightest particles are excited, while at higher temperatures, heavier particles take over.
The massive particles are largely unaffected by the presence of magnons, leading to a continuous momentum Onsager matrix as a function of coupling.

\section{High-temperature limit and divergence of charge Onsager matrix} \label{sec:highT}

\begin{figure}
    \centering
    \includegraphics[width=1.0\columnwidth]{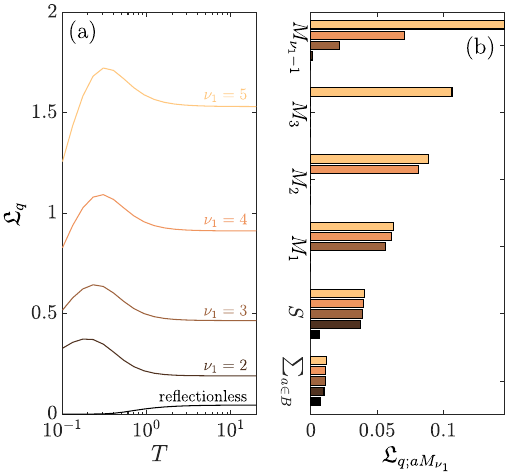}
    \phantomsubfloat{\label{fig:highT_plateau_a}}
    \phantomsubfloat{\label{fig:highT_plateau_b}}
    \vspace{-2\baselineskip}% Remove extra line inserted by subfloat
    
    \caption{(a) Topological charge Onsager matrix~\eqref{eq:Onsager} as function of temperature $T$ for couplings with increasing number of level-1 magnon species $\nu_1$. Calculated for $N_B = 3$ breathers. (b) High-temperature limit of contributions to the Onsager matrix from scattering of the last magnon scattering with remaining quasi-particles.}
    \label{fig:highT_plateau}
\end{figure}

At high temperatures, the interaction term in the sine--Gordon model can typically be neglected, whereby the system approximates a massless free boson. This simplification was used to derive the high-temperature limit of charge Drude weight~\cite{2023PhRvB.108x1105N}. 
Since diffusion is driven by two-particle interactions, one would, therefore, expect the Onsager matrix to vanish in the high-temperature limit. 
However, when calculating the charge Onsager matrix for increasing temperatures $T$, we find that it instead converges to a finite value, as shown in Fig.~\ref{fig:highT_plateau_a}.
The free boson theory, which only accounts for ultra-relativistic particles, cannot explain this high-temperature convergence.
Instead, the Onsager contribution must come from low-energy particles:
In the limit $T \gg m_a$, massive modes are highly excited. However, their corresponding density of states remains small at lower rapidities.
Thus, particles are excited at higher rapidities, leading to root densities $\rho_a (\theta)$ peaked around $\theta \sim \pm \log 2T/m_a$ and otherwise exponentially close to zero.
As a result, the dressing equations decouple into independent left and right moving modes, and excitations around the ultra-relativistic peaks have effective velocity $\pm 1$.
From the Onsager matrix expression~\eqref{eq:Onsager}, the term $\left| v_a^{\text{eff}}(\theta_1) - v_b^{\text{eff}}(\theta_2) \right|$ shows that scattering between particles with the same effective velocity, i.e.~within the same rapidity peak, does not contribute.
Further, the dressed scattering kernel $\Phi_{a b}^{\mathrm{dr}} (\theta_1 - \theta_2)$ rapidly vanishes for large rapidity differences, meaning scattering between particles in opposite peaks is also negligible.
Therefore, only the scattering of states $\theta \sim 0$, which remain constant beyond a certain temperature, significantly contributes to the charge Onsager matrix, thus leading to its convergence.

\begin{figure}
    \centering
    \includegraphics[width=1.0\columnwidth]{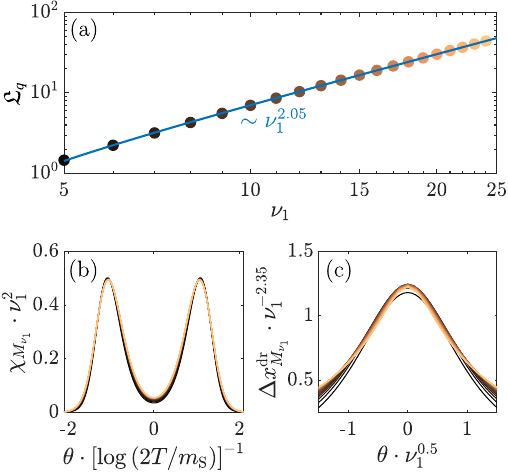}
    \phantomsubfloat{\label{fig:highT_Onsager_scaling_a}}
    \phantomsubfloat{\label{fig:highT_Onsager_scaling_b}}
    \phantomsubfloat{\label{fig:highT_Onsager_scaling_c}}
    \vspace{-2\baselineskip}% Remove extra line inserted by subfloat
    
    \caption{Divergence of charge conductivity and magnon scattering at high temperature, calculated for $T = 20$ and $N_B = 1$.
    (a) Charge Onsager matrix~\eqref{eq:Onsager} as function of number of level-1 magnons $\nu_1$. The observed divergence is polynomial $\mathfrak{L}_q \sim \nu_1^\gamma$, demonstrated by the fitted blue line.
    (b) Susceptibility of the last magnon. 
    (c) Dressed displacement of the last magnon when scattering with the second-to-last magnon. At low rapidities, a reasonable collapse of the curves is found following appropriate scaling. 
    The color of each curve corresponds to the points in (a).}
    \label{fig:highT_plateau}
\end{figure}

In the high-temperature limit, the charge Onsager matrix diverges with an increasing number of magnons.
Following dressing, only the last two magnons carry a topological charge of value $\pm \nu_1$.
Hence, we can restrict our study of the Onsager divergence to scattering processes involving the last magnon, which we plot in Fig.~\ref{fig:highT_plateau_b}.
For $\nu_1 \gg 1$, we find that scattering between the two last magnons produces the dominant contribution.

To study this scattering process in more detail, we calculate the high-temperature charge Onsager matrix for up to 24 magnon species; the results are very accurately fitted with the power-law $\mathfrak{L}_q \sim \nu_1^{\gamma}$ with $\gamma \approx 2.05$, as shown in Fig.~\ref{fig:highT_Onsager_scaling_a}.
Note that all exponents reported here depend weakly on the number of breathers.
The contribution to the charge Onsager matrix from the scattering between the last two magnon species exhibits the same scaling.
A similar power-law divergence is seen in the d.c. spin conductivity of the XXZ spin chain at infinite temperature near irrational coupling~\cite{2020PhRvB.101v4415A}.
There, the dominant scattering events experienced by charged particles are those with the heaviest neutral quasi-particles.
These heavy particles become rarer when approaching an irrational coupling. However, the associated Wigner scattering displacement of the charged particle exhibits a power-law increase.
Thus, the charged quasi-particle undergoes Levy flights.
For the sine--Gordon model, the magnons are auxiliary excitations accounting for internal degeneracies of the charge degrees of freedom, thus complicating a physical interpretation of the dynamics of the charged particles.
However, we can still analyze the dynamics of the magnons using the soliton gas framework.

In Fig.~\ref{fig:highT_Onsager_scaling_b}, we plot the susceptibilities $\chi$ of the last magnon.
For rapidities near the ultrarelativistic peaks $\theta \sim \pm \log 2T/m_\mathrm{S}$, we find an inverse quadratic scaling; for low rapidity states $\theta \sim 0$, which are the main contributor to the Onsager matrix at high temperature, the scaling is slightly different $\chi_{M_{\nu_1}} (0) \sim \nu_1^{-1.8}$.
Next, in Fig.~\ref{fig:highT_Onsager_scaling_c}, we plot the dressed displacement of the last magnon when scattering with the second-to-last magnon.
The peak displacement at $\theta = 0$ exhibits an approximate power-law divergence scaling as $\Delta x^{\mathrm{dr}} (0) =  \Phi^{\mathrm{dr}} / (\partial_\theta p)^{\mathrm{dr}} (0) \sim \nu_1 ^{2.35}$.
These scaling behaviors are similar to those observed in the XXZ chain~\cite{2020PhRvB.101v4415A} (though for a different coupling parameter), thus suggesting that the anomalous charge conductivity of sine--Gordon model likewise is due to the charged quasi-particles undergoing Levy flights.

\section{Low-temperature limit} \label{sec:lowT}

\begin{figure}
    \centering
    \includegraphics[width=1.0\columnwidth]{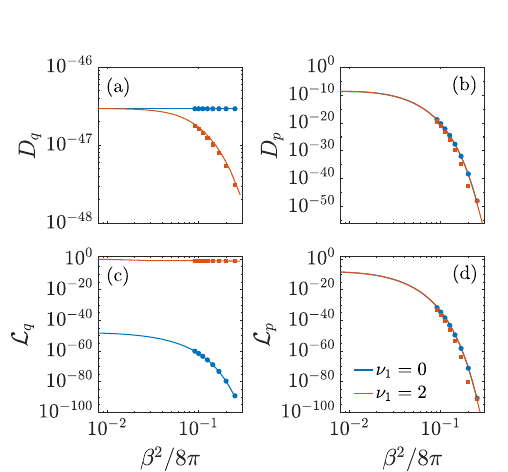}
    \phantomsubfloat{\label{fig:lowT_approximation_a}}
    \phantomsubfloat{\label{fig:lowT_approximation_b}}
    \phantomsubfloat{\label{fig:lowT_approximation_c}}
    \phantomsubfloat{\label{fig:lowT_approximation_d}}
    \vspace{-2\baselineskip}% Remove extra line inserted by subfloat
    
    \caption{Transport coefficients as function of coupling at low temperature ($T = 0.01$, close to the gap at the lowest coupling).
    The blue and red curves are calculated for reflectionless and attractive 2-magnon points using the simplified low-temperature TBA equations, accounting only for select quasi-particle scattering processes.
    Symbols in corresponding colors show the results of full GHD calculations.}
    \label{fig:lowT_approximation}
\end{figure}

The analysis of transport conducted in Ref.~\cite{2019PhRvB.100c5108B} showed that semi-classical predictions of the sine--Gordon model are analytically recovered from GHD in the low-temperature limit. However, this treatment was restricted to the repulsive regime $\xi > 1$.
To extend the analysis to any coupling, we take advantage of the vanishing population of all but the lightest massive particles in the low-temperature limit, which, as already seen in  Fig.~\ref{fig:Onsager_particlewise}, drastically reduces the relevant scattering contributions to the Onsager matrix.
Further, the inter-magnon scattering vanishes for $\nu_1 = 2$, leaving only the soliton-magnon contribution.
Thus, the scattering kernel takes a particularly simple form, allowing us to efficiently compute transport coefficients (see Appendix for expressions).
Owing to the simplicity of the low-temperature formulation, we can extend the calculations to several hundreds of breather species.
Approaching the $\beta \to 0$ limit is of particular interest, as experimental realizations of the sine--Gordon model achieved with tunnel-coupled quasi-condensates are in the vicinity of this regime~\cite{2017Natur.545..323S, 2018PhRvL.120q3601P, 2021NatPh..17..559S, 2023NatPh..19.1022T}.

In Fig.~\ref{fig:lowT_approximation}, we show the results of calculating the transport coefficients of charge and momentum at $T = 0.01$ for the reflectionless and two-magnon points.
To test our low-temperature approximation, we compare it with GHD computations employing the full scattering kernel; for couplings with few breather types, we observe a good agreement.
For a fixed temperature $T$, the density of solitons (and kinks/antikinks) $\rho_S (\theta)$ remains constant when lowering the coupling, while the density of magnons changes slightly (converges for sufficiently low $\beta$).
Meanwhile, since the mass of the first breather $m_{B_1} = 2 m_S \sin(\pi \xi /2)$ decreases with coupling, the density of breathers increases significantly as $\beta$ is lowered.
This results in a growth of transport coefficients containing breather contributions, as seen in Fig.~\ref{fig:lowT_approximation}.

When increasingly many breather species enter, their spectrum effectively becomes continuous, and the sine--Gordon model becomes classical with $\lim_{\xi \to 0} S_R /S_T = 0$.
Thus, all anomalous transport phenomena related to reflective scattering should vanish in the limit $\xi \to 0$, and transport coefficients should be well-represented by their values at the reflectionless points.
This classical limit has been obtained through the reflectionless points~\cite{koch2023exact, 2024PhRvB.109c5118B}.
Indeed, for sufficiently low coupling strengths, the charge Drude weights $D_q$ computed for the reflectionless and attractive points converge to the same value, whereby the fractal structure vanishes.
However, we find that $\mathfrak{L}_q$ of reflectionless and magnonic points \textit{do not} converge; instead, the charge diffusion of the magnonic points continues to be many orders of magnitude greater.
We attribute this to a conflict of limits: Upon approaching the classical limit, the amplitude of reflective scattering should vanish. However, by calculating the transport coefficient using the scattering of magnons, we implicitly assume that $S_R$ is finite. Thus, the classical result can only be obtained by taking the limit $\xi \to 0$ through reflectionless points.

\section{Crossover time scale and dynamics at finite scales} \label{sec:finite_scales}

The continuity equations $\partial_t n_i + \partial_x j_i = 0$ link the Drude weight and Onsager matrix to the spreading of a local charge perturbation, whose density correlation function at asymptotic long times reads~\cite{2019ScPP....6...49D}
\begin{equation}
    \int \text{d}x\, x^2 \langle n_i(x,t) n_i(0,0) \rangle^c = D_{i}t^2 + \mathfrak{L}_{i} t + \mathcal{O}(t)\,.
    \label{eq:DrudeOnsager}
\end{equation}
To the leading order, correlations spread ballistically at a rate governed by the Drude weights. At the same time, the term $\mathfrak{L}_{i} t$ represents diffusive broadening around the ballistic propagation quantified by the Onsager matrix.
Hence, while transport is ultimately ballistic, charges may spread diffusively at short to intermediate time scales, particularly if the corresponding Onsager matrix is large.
Notably, experiments are limited to probing only shorter time scales.

We therefore introduce a crossover time-scale between diffusive and ballistic hydrodynamic transport $t_i^{\star}$, defined as the ratio
\begin{equation}
    t_i^{\star} = \frac{\mathfrak{L}_i}{D_i} \; .
\end{equation}
From the previous calculations of transport coefficients shown in Fig.~\ref{fig:transport_coefficients}, we find that momentum transport is dominantly ballistic, even at short time scales, as $ t_p^{\star} \ll 1$.
Other integrable models with diagonal scattering, such as the one-dimensional Bose gas and the reflectionless points of the sine--Gordon model, display similarly weak diffusion~\cite{2024ScPC....7...25M}.
Meanwhile, at generic couplings of the sine--Gordon model, the nature of charge transport is dependent on temperature:
Following our analysis of the low-temperature limit $T \ll m_S$ in Sec.~\ref{sec:lowT}, it is evident that the charge spreads diffusively at all but the very longest time scales.
Oppositely, in the high-temperature limit $T \gg m_S$, we find that the charge Onsager matrix saturates, whereas the charge Drude weight increases linearly with $T$~\cite{2023PhRvB.108x1105N}. 
Hence, competition ensues between the temperature scaling of the Drude weight and the divergence of the Onsager matrix with increasing magnon species.
At intermediate temperatures, $T \sim m_S$, we find that $t_{q}^{\star} \sim 1$ for attractive points with two magnon species; increasing the number of magnon species again leads to a divergence of the charge Onsager matrix, and thus the crossover time-scale.

\begin{figure}
    \centering
    \includegraphics[width=1.0\columnwidth]{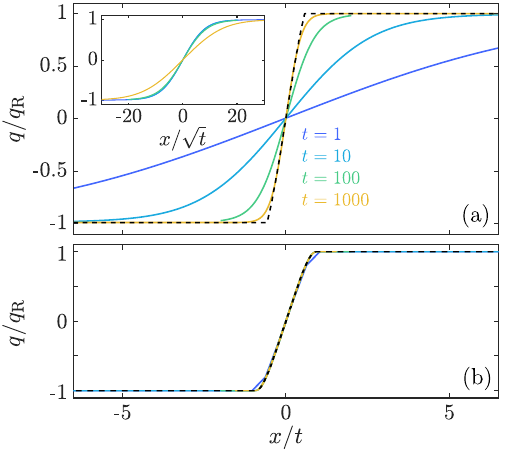}
    \phantomsubfloat{\label{fig:bipartition_a}}
    \phantomsubfloat{\label{fig:bipartition_b}}
    \vspace{-2\baselineskip}% Remove extra line inserted by subfloat
    
    \caption{Topological charge density at times $t$ following a quench from a bipartite thermal state at temperature $T = 0.25$ and chemical potential $\mu_{\mathrm{L}/\mathrm{R}} = \pm 0.001$. (a) Attractive point with $N_B = 1$ breathers and $\nu_1 = 2$ magnons, resulting in a crossover time of $t_{q}^\star = 517$. The solid lines correspond to solutions of the diffusive equation~\eqref{eq:bipartition_diffusive}, while the dashed line is the ballistic solution~\eqref{eq:bipartition_ballistic}. The inset shows a collapse of the solutions at shorter time scales following diffusive rescaling. (b) Reflectionless point with $N_B = 2$ breathers and $t_{q}^\star = 0.06$.}
    \label{fig:bipartition}
\end{figure}

To demonstrate the crossover between diffusive and ballistic hydrodynamics, consider the bipartition protocol, where the system is initialized with the step condition $ \vartheta_a(\theta ; x, 0) = \vartheta_{a;\mathrm{L}}(\theta) \Theta(-x) + \vartheta_{a;\mathrm{R}}(\theta) \Theta(x)$, with $\vartheta_a = \rho_a / \rho_a^{\mathrm{tot}}$ and $\Theta(x)$ being the Heaviside function.
Given just a small difference between the left and right subsystems, $\vartheta_{a;\mathrm{L}} \sim \vartheta_{a;\mathrm{R}}$, the Euler-scale solution at time $t$ is given by
\begin{equation}
    \vartheta_a(\theta ; x, t) = \vartheta_{a;\mathrm{L}}(\theta) - \left(\vartheta_{a;\mathrm{R}}(\theta) - \vartheta_{a;\mathrm{L}}(\theta)\right) \Theta\left(t v_a^{\text {eff}}(\theta)-x\right) \; .
    \label{eq:bipartition_ballistic}
\end{equation}
For each fluid mode (or rapidity $\theta$), the initial interface between the initial subsystems translates to the coordinate fulfilling $v_a^{\text{eff}}(\theta) = x/t$.
The resulting ballistic currents are determined by the Drude weights~\cite{2017PhRvL.119b0602I}; thus, the bipartition setup can be used to measure $D_i$~\cite{arXiv:2406.17569}.

In the presence of diffusion, the boundary is smoothened out, and the solution up to corrections of order $t^{-1}$ instead reads~\cite{2019ScPP....6...49D}
\begin{equation}
\begin{aligned}
    \vartheta_a(\theta ; x, t) =& \: \vartheta_{a;\mathrm{L}}(\theta)+ \frac{\left(\vartheta_{a;\mathrm{R}}(\theta) + \vartheta_{a;\mathrm{L}}(\theta)\right)}{2} \times \\
    &\times \left( 1-\operatorname{erf}\left[\sqrt{\frac{t v_a^{\mathrm{eff}}(\theta)-x}{4 t w_a(\theta)}} \right] \right) + \mathcal{O}(t^{-1})
\end{aligned}
\label{eq:bipartition_diffusive}
\end{equation}
where $w_a (\theta)$, which quantifies the diffusive broadening of a quasi-particle of species $a$ and rapidity $\theta$~\cite{2018PhRvB..98v0303G}, is given by
\begin{equation}
\begin{aligned}
    w_a(\theta) =& \sum_b \int \mathrm{~d} \theta^{\prime} \; \bigg[ \chi_b(\theta ') \left(\frac{\Phi_{a b}^{\mathrm{dr}}\left(\theta - \theta^{\prime}\right) \eta_b}{2 \pi \rho_a^{\text{tot}}(\theta)}\right)^2 \times \\ 
    &\qquad \times \left|v_a^{\mathrm{eff}}(\theta) - v_b^{\mathrm{eff}}\left(\theta^{\prime}\right) \right| \bigg] \; .
\end{aligned}
\end{equation}
Note, in Eq.~\eqref{eq:bipartition_diffusive}, we have omitted an additional term describing rearrangements of rapidities among particles following inter-particle scatterings, as this term only yields a minor contribution to charge transport.

Given a bipartite thermal state with small, opposite polarization, the charge profiles of the solutions to Eqs.~\eqref{eq:bipartition_ballistic} and \eqref{eq:bipartition_diffusive} for a generic attractive as well as a reflectionless point are shown in Fig.~\ref{fig:bipartition}.
For the reflectionless point, the solutions to both equations collapse to a single curve following ballistic rescaling, consistent with the very short corresponding crossover time-scale of $t_{q}^\star = 0.06$.
In the presence of reflective scattering, however, we find that charge profiles for $t \ll t_{q}^\star = 517$ follow diffusion scaling, whereas later solutions at $t \gg t_{q}^\star$ approach the ballistic curve.
This may explain the observations in Ref.~\cite{2019PhRvB.100c5108B}, where a hybrid approach, combining classical quasi-particle trajectories with a quantum description of kink-antikink scattering~\cite{2017PhRvL.119j0603M}, was used to study charge dynamics.
After a bipartite quench, the charge front showed diffusive scaling, which seemed at odds with the ballistic transport predicted by GHD.
However, given the very long diffusive time scale for non-reflectionless points, it is likely that the asymptotic ballistic regime was not reached within the evolution times explored.

\begin{figure}
    \centering
    \includegraphics[width=1.0\columnwidth]{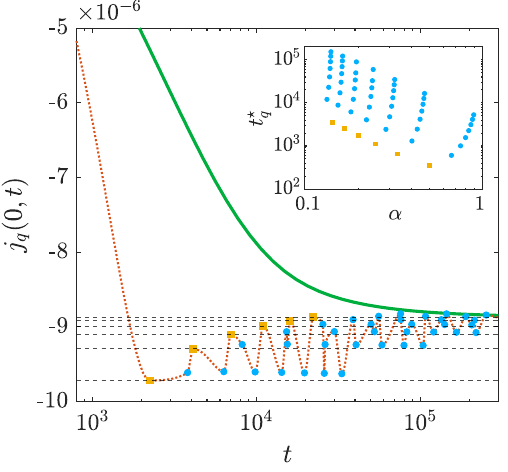}
    \caption{Hydrodynamic charge current at $x=0$ following a quench of a bipartite state (green curve) for $\xi = \left( 2 + 1/7 \right)^{-1}$. Markers indicate $ 2 \pi t_{q}^\star$ and ballistic currents calculated from the solution to Eq.~\eqref{eq:bipartition_ballistic} for couplings with an increasing number of level 1 (orange squares) and 2 magnons (blue circles) until $\xi$ is reached. The red dotted curve connecting the points is a guide to the eye. Dashed lines indicate the asymptotic currents (plateaus) of couplings $\xi =  \left( 2 + 1/\nu_1 \right)^{-1}$ for $\nu_1 = 2 , \ldots , 7$. The inset shows the corresponding crossover times $t_{q}^\star$ as functions of the continued fraction $\alpha = (  \nu_1 + ( \nu_2 + \ldots )^{-1} )^{-1}$ for $N_B = 2$ breathers.}
    \label{fig:transient_current}
\end{figure}

Approaching a reflectionless point, the charge Onsager matrix, and thus $t_q^{\star}$, diverges, suggesting that charge transport becomes fully diffusive.
However, microscopic simulations of XXZ spin chains near irrational coupling strengths suggest that dynamics at finite time scales is quite different~\cite{2018PhRvB..97h1111C}.
Following a quench, the induced current passes through transient regimes, exhibiting plateaus corresponding to the Drude weights of subsequent rational approximations of the (irrational) coupling.
Physical intuition of this phenomenon is given by the quasi-particle spectrum, which in the XXZ spin chain consists of neutral particles of increasingly larger size; upon approaching the irrational coupling point, the number of particle species grows, similar to the quantum sine--Gordon model.
At finite evolution times, a propagating charge-carrying particle can, therefore, only resolve part of the excitation spectrum up to particles of a certain length.
Thus, the system effectively acquires transient transport coefficients corresponding to the subsequent rational approximations of the coupling.
Notably, this phenomenon is completely absent in GHD, as the diffusion kernel is derived from the quasi-particle distributions of the thermodynamic Bethe Ansatz; being formulated in the thermodynamic limit, the theory immediately resolves the full particle spectrum, whereby the GHD dynamics exhibits no transient behavior.
Indeed, as shown in Fig.~\ref{fig:transient_current} for $\xi = \left( 2 + 1/7 \right)^{-1}$ in the quantum sine--Gordon model, calculating the charge current of the hydrodynamic bipartition solution~\eqref{eq:bipartition_diffusive} yields a smooth crossover to the asymptotic value set by the Drude weight. 
However, following the close similarities of the Bethe Ansatz solution between the XXZ spin chain and the quantum sine--Gordon model, it is likely that a charge current induced in the latter would also exhibit transient behavior. 

To illustrate how the resulting charge current may appear, we compute the asymptotic currents and crossover time-scales for couplings approaching $\xi = 1/2$, i.e.~for two breather species with an increasing number of magnon species.
The results, connected by dotted curves to guide the eye, are plotted in Fig.~\ref{fig:transient_current}, where plateaus corresponding to increments of level-1 magnons are highlighted by dashed lines.
Moving through couplings with increasing crossover time, we find an oscillatory behavior of the ballistic charge current while increasing $\nu_1$ leads to an increase in $t_{q}^\star$ and a decrease in $D_q$. Accounting for level-2 magnonic excitations introduces intermediate crossover times. Additional magnon levels likely yield even finer features. However, achieving numerical convergence of the TBA equations for larger numbers of magnon species becomes increasingly difficult.

\section{Conclusions}\label{sec:conclusions}

In this work, we have comprehensively studied transport properties in the integrable quantum sine--Gordon model.
Specifically, we have focused on the transport of topological charge, which shows anomalous behavior in the low-frequency regime due to non-diagonal scattering between the physical charge-carrying particles.
At reflectionless points, where scattering becomes diagonal for specific coupling values, charge transport is ballistic, while diffusive effects merely provide subleading contributions to dynamics.
However, for generic couplings, diffusive transport dominates at most accessible time scales, unlike the behavior found in most integrable models.

As the coupling approaches a reflectionless point, the charge Onsager matrix diverges with power-law scaling, causing the diffusive time scale to increase.
To identify the source of this diffusion, we have analyzed scattering between quasi-particle excitations and found that the primary contribution comes from interactions between charged particles and those carrying \textit{dressed} charge.
By reducing the Onsager matrix to a few key scattering combinations, we could efficiently examine transport in both low- and high-temperature limits.

We also observed that the scattering of dressed charge carriers, particularly the two last magnon types, shows similar scaling with coupling strength, as seen in the integrable XXZ spin-1/2 chain.
In that system, the power-law divergence of the Onsager matrix was linked to Levy flights of spin-carrying excitations.
We anticipate a similar mechanism to be responsible for the charge Onsager matrix divergence of the sine--Gordon model since solving its non-diagonal scattering problem requires auxiliary Bethe equations identical to those of the gapless phase of the XXZ spin chain.

Analogously with the XXZ spin chain, at finite times and in finite-sized systems, the sine--Gordon model's transport dynamics is likely to exhibit transient behavior as its extensive excitation spectrum is dynamically resolved.
However, because Generalized Hydrodynamics (GHD) is formulated in the thermodynamic limit, capturing all excitation species at all times, simulations of microscopic dynamics are needed to observe such transient effects.

GHD and the Bethe Ansatz quasi-particle framework offer new perspectives on many-body dynamics, particularly in analog quantum field simulators realized using ultracold atoms.
For example, studies of the integrable one-dimensional Bose gas have deepened our understanding of the emergence and breakdown of effective field theories.
Recent advancements in theoretical techniques have provided valuable insights into sine--Gordon dynamics~\cite{2018PhRvL.121k0402K, 2019PhRvA.100a3613H, 2022ScPP...12..144H}; to achieve a comprehensive understanding of the model, it is thus crucial to consolidate these findings with the GHD framework.
A key initial step in this direction is exploring the transition between the quantum and classical sine--Gordon models, along with their respective GHD formulations~\cite{koch2023exact, 2024PhRvB.109c5118B, DelVecchio2023, 2024ScPP...16..145N}, since current experimental quantum field simulators operate in this transition regime.

\begin{acknowledgments}

We thank Jacopo de Nardis for useful discussions.
This work was supported by the National Research, Development and Innovation Office (NKFIH) through the OTKA Grant ANN 142584. FM acknowledges support from the European Research Council: ERC-AdG: Emergence in Quantum Physics (EmQ) and partial support by the Austrian Science Fund (FWF) (Grant No. I6276). BN was also partially Supported by the EKÖP-24-3-BME-62 University Research Fellowship Programme of the Ministry for Culture and Innovation from the source of the National Research, Development and Innovation Fund of the Ministry of Culture and Innovation and the Budapest University of Technology and Economics under a grant agreement with the National Research, Development and Innovation Office. GT was partially supported by the Quantum Information National Laboratory of Hungary (Grant No. 2022-2.1.1-NL-2022-00004).
\end{acknowledgments}

\bibliographystyle{utphys}
\bibliography{anomalous_transport}
%\nocite{*}

%\clearpage

\appendix
\section{Generalized hydrodynamics for the sine--Gordon model}\label{sec:sG_GHD}

\subsection{Sine--Gordon thermodynamics}\label{subsec:sG_thermodynamics}

The main complication in setting up sine--Gordon thermodynamics and hydrodynamics is the non-diagonal character of the kink-antikink scattering 
\begin{align}
&S_{++}^{++}(\theta)=S_{--}^{--}(\theta)=S_0(\theta)\,,\nonumber\\
&S_{+-}^{+-}(\theta)=S_{-+}^{-+}(\theta)=S_T(\theta)S_0(\theta)\,,\nonumber\\
&S_{+-}^{-+}(\theta)=S_{-+}^{+-}(\theta)=S_R(\theta)S_0(\theta)\,,\nonumber\\
&S_T(\theta) = \frac{\sinh\left(\frac{\theta}{\xi}\right)}{\sinh\left(\frac{i\pi-\theta}{\xi}\right)}\,,\quad 
S_R(\theta) = \frac{i\sin\left(\frac{\pi}{\xi}\right)}{\sinh\left(\frac{i\pi-\theta}{\xi}\right)}\,,\nonumber\\
&S_0(\theta) = -\exp\left(i\int\limits_{-\infty}^{\infty} \frac{\mathrm{d}t}{t}
                  \frac{\sinh\left(\frac{t\pi}{2}(\xi-1)\right)}{2\sinh\left(\frac{\pi\xi t}{2}\right)\cosh\left(\frac{\pi t}{2}\right)}{e}^{i\theta t}\right)\,,
\label{SS_scattering_matrix}
\end{align}
where $+/-$ stands for kinks/antikinks, with $\theta$ denoting the difference of their rapidities. $S_T$ and $S_R$ are the amplitudes for transmission and reflection, with integer values of $1/\xi$ corresponding to reflectionless points. The scattering theory can be diagonalised using magnonic excitations, as detailed in Subsection \ref{subsec:TBA}. 

In the thermodynamic limit, the system can be described in terms of the total densities of states $\rho_a^{\text{tot}}$ and densities of occupied states $\rho_a$ of quasi-particle excitation $a$, with the Bethe Ansatz relating them by the following system of linear integral equations: 
\begin{equation}
    \rho_a^{\text{tot}} = \rho_a+\rho_a^{(h)} = \eta_a \frac{m_a}{2\pi}\cosh\theta + \sum_b \eta_a \Phi_{ab}*\rho_b \ ,
\label{eq:BA_with_densities}
\end{equation}
where $m_a=0$ for magnonic degrees of freedom, $\rho_a^{(h)}$ is the density of unoccupied states (holes), and the star denotes convolution in rapidity space defined by
\begin{equation}
   (f*g)(\theta) = \int \frac{\text{d}\theta'}{2\pi} f(\theta-\theta')g(\theta')\,.
   \label{eq:convolution}
\end{equation}
The $\Phi_{ab}$ kernels are the logarithmic derivatives of the scattering phases between the different quasi-particles, while $\eta_a=\pm 1$ are appropriate sign factors; for their explicit specification, we refer to \cite{2024ScPP...16..145N}. These equations can be brought to a partially decoupled form, which has several advantages: it has simpler kernels given by universal functions, can be encoded graphically and speeds up the numerical solution of the system \cite{2023PhRvB.108x1105N,2024ScPP...16..145N}.

Equations \eqref{eq:BA_with_densities} are valid for any equilibrium state of the system. Integrable systems have more general equilibrium states due to the presence of infinitely many conserved charges, which are often conceptualized as generalized Gibbs ensembles \cite{2016JSMTE..06.4007V}. However, it turned out that the local conserved charges were generally insufficient to describe such equilibrium states \cite{2014PhRvL.113k7203P,2014PhRvL.113k7202W}, and they must be completed using quasi-local charges \cite{2015PhRvL.115o7201I}. As it turns out, the specification of the equilibrium state in terms of the complete set of charges is equivalent to giving all the densities of occupied states $\rho_a$ via the correspondence known as string-charge relations \cite{Ilievski:2015nqt,Feher:2019naf}. Therefore, it is preferable to consider the general equilibrium state to be described by the densities, avoiding the problems with finding the complete set of charges. Nevertheless, we note that such charges have been found for the sine--Gordon model as well \cite{Vernier:2016qcc}.

To find a specific equilibrium state, additional conditions beyond \eqref{eq:BA_with_densities} are required. For the important case of thermal equilibrium at temperature $T$ and chemical potential $\mu$ (coupled to the topological charge), these are given by the following thermodynamic Bethe Ansatz (TBA) equations:
\begin{equation}
    \epsilon_a = w_a - \sum_b \eta_b \Phi_{ab}*\log(1+\text{e}^{-\epsilon_b})\,,
\label{eq:TBA}\end{equation}
where the unknowns to be solved for are the pseudo-energy functions
\begin{equation}
    \epsilon_a=\log\frac{\rho_a^{(h)}}{\rho_a}\,,
\end{equation}
and 
\begin{equation}
    w_a=\frac{m_a}{T} \cosh\theta-\frac{\mu q_a}{T}
\end{equation}
correspond to the bare quasi-particle energies where $q_a$ gives the topological charge of the excitation species $a$.

\begin{figure}
    \centering
    \includegraphics[width=0.45\linewidth]{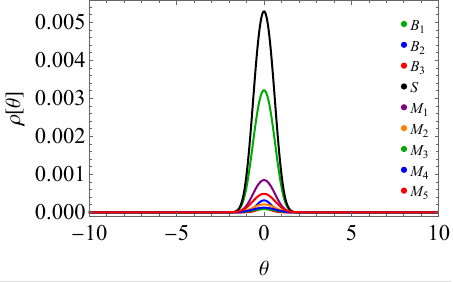}
    \hfil
    \includegraphics[width=0.45\linewidth]{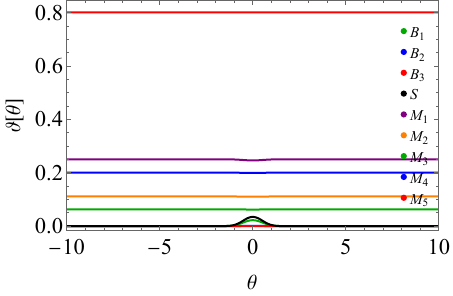}
    \includegraphics[width=0.45\linewidth]{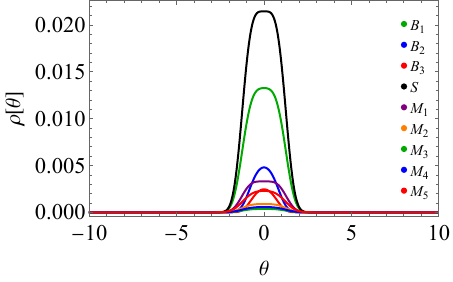}
    \hfil
    \includegraphics[width=0.45\linewidth]{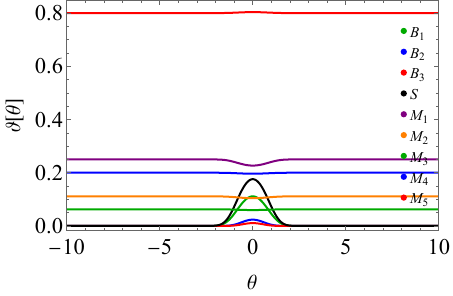}
    \includegraphics[width=0.45\linewidth]{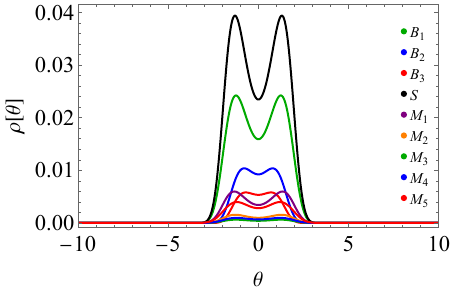}
    \hfil
    \includegraphics[width=0.45\linewidth]{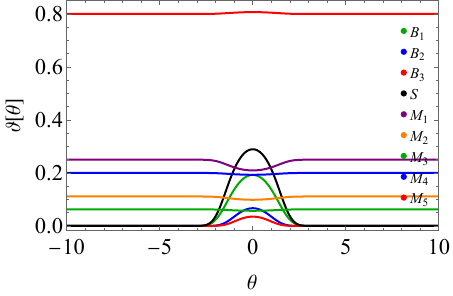}
    \includegraphics[width=0.45\linewidth]{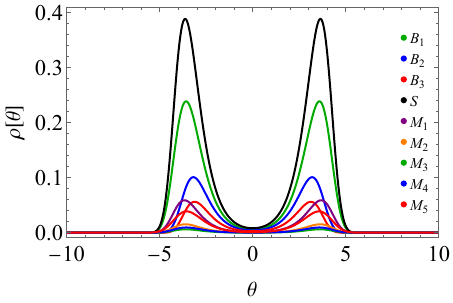}
    \hfil
    \includegraphics[width=0.45\linewidth]{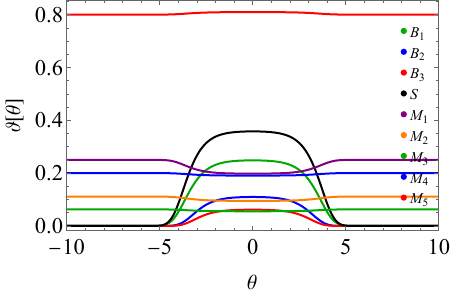}
    \includegraphics[width=0.45\linewidth]{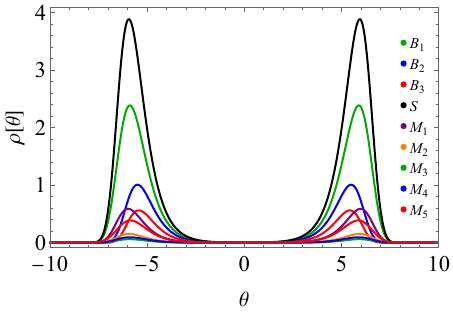}
    \hfil
    \includegraphics[width=0.45\linewidth]{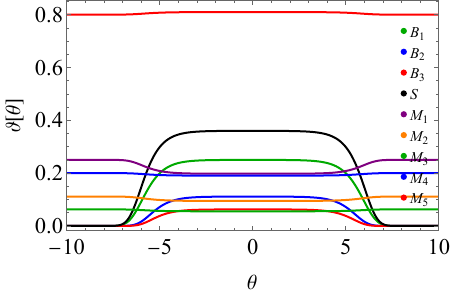}
    \caption{Example particle density and filling profiles for $T=0.25$, $0.5$, $1$, $10$, $100$ from top to bottom.}
    \label{fig:ex_den_fill}
\end{figure}

Figure \ref{fig:ex_den_fill} shows examples of density profiles of occupied states $\rho_a(\theta)$ and filling functions $\vartheta_a(\theta)$, highlighting the emergence of the double-peaked distribution of particles for high temperatures. Note also that the filling functions of magnons take finite asymptotic values for $|\theta|\to \infty$. Unlike massive particles, the magnons' total densities of states (not shown) vanish as $|\theta|\to\infty$, yielding a finite number of occupied states.

\subsection{Mapping between the sine--Gordon and the XXZ TBA}

The Hamiltonian of the XXZ spin chain reads
\begin{equation}
    H_{\text{XXZ}} = -\sum_{i} S_i^x S_{i+1}^x + S_i^y S_{i+1}^y + \Delta S_i^z S_{i+1}^z\,.
\end{equation}
In case $-1<\Delta<1$ (which has a quasiparticle content different from the $\left|\Delta\right|>1$ case), the Bethe Ansatz equations of the model \cite{takahashi_1999} can be mapped to that of the magnonic equations of the sine--Gordon model for reflective couplings. The coupling strength is usually parametrized as $\Delta = \cos\gamma$ with $0<\gamma<\pi$, and the identification of the XXZ parameter in terms of the sine-Gordon coupling is given by
\begin{equation}
    \gamma=\frac{\pi}{\alpha}
    \hspace{0.5cm}
    \text{with}
    \hspace{0.5cm}
    \xi = \frac{1}{N_B+\frac{1}{\alpha}}\,.
\end{equation}
Writing $\alpha$ in terms of its (unique) continued fraction
\begin{equation}
    \alpha = \nu_1+\frac{1}{\nu_2+\dots}
\end{equation}
determines the magnon content in the TBA equations, which consists of $\nu_i$ level-$i$ magnons, i.e., $\sum_i \nu_i$ magnons altogether.

\begin{figure}
    \centering
    \includegraphics[width=1\linewidth]{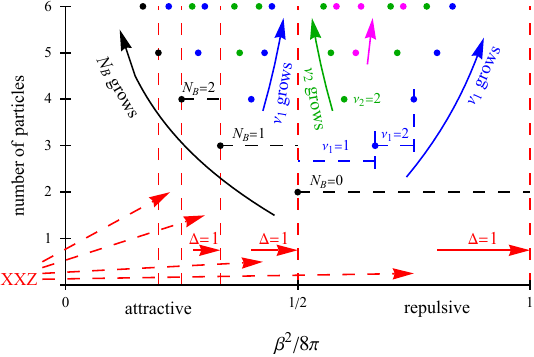}
    \caption{TBA particle content and mapping of the sine--Gordon model to the XXZ spin chain.}
    \label{sG_XXZ_map}
\end{figure}

The mapping works in both the repulsive ($N_B=0$) and the attractive ($N_B\neq 0$) regimes by appropriately shifting and rescaling the rapidities (see \cite{2024ScPP...16..145N} for details). Still, it is otherwise independent of the number of breathers. This means that couplings for which $\alpha$ is the same but $N_B$ is different have the same internal XXZ Bethe Ansatz as shown in Figure \ref{sG_XXZ_map}.

In the XXZ model, when $\Delta=1$ (the XXX model), both the TBA equations and the structure of magnons differ from the cases where $|\Delta|<1$. In the repulsive regime, only the point $\beta^2/8\pi=1$ corresponds to $\Delta=1$. At this point, it is possible to relate the string structure of the XXX model to the magnon structure of the sine--Gordon model.

In the attractive regime, the only points with $|\Delta|=1$ are the ``reflectionless points" where $1/\xi=N_B+1$, resulting in $\alpha=1$ and $\Delta=\cos \pi/\alpha =-1$. However, at exactly the reflectionless points, the Bethe Ansatz equations contain no nesting, so there is no corresponding spin chain. As a result, the sine--Gordon TBA excitation spectrum contains no magnons; instead, it solely consists of $N_B=1/\xi-1$ species of breathers and a soliton/anti-soliton pair, all of which scatter diagonally \cite{ZAMOLODCHIKOV1991391}.

\subsection{Generalized hydrodynamics for the sine--Gordon model}\label{subsec:sG_hydro}

On the hydrodynamic (Euler) scale, the dynamics of integrable systems is captured by the theory of Generalized Hydrodynamics. It is based on the transport of the infinitely many conserved quantities expressed by the continuity equation
\begin{equation}
    \partial_t n_i(x,t)+ \partial_x j_i (x,t) = 0\,,
\end{equation}
where the expectation value of the local charges is given by 
\begin{equation}
\label{expvalues_q}
        n_i = \sum_a \int \mathrm{d}\theta \rho_a(\theta)  h_{i;a}(\theta)  %=\sum_a \int \frac{\mathrm{d}\theta}{2\pi} m_a\cosh\theta\,\vartheta_a(\theta) h_a^{\textrm{dr}}(\theta)
        \,.
\end{equation}  
The expression of the current densities, conjectured in \cite{2016PhRvX...6d1065C,2016PhRvL.117t7201B,2017ScPP....2...14D} and then rigorously established in \cite{2020PhRvX..10a1054B,2020ScPP....8...16P,2020PhRvL.125g0602P} reads
\begin{equation}
        j_{i} = \sum_a \int \mathrm{d}\theta \rho_a(\theta) h_{i;a}(\theta) v_a^{\textrm{eff}}(\theta)\,,\label{eq:jh}
\end{equation}
where
% \begin{equation}
%     \vartheta_i(\theta)=\frac{1}{1+e^\epsilon_i(\theta)}
% \end{equation}
% is the filling fraction and 
the effective velocity $v_a^{\textrm{eff}}$ accounts for the accumulated contributions of scattering delays as described in Subsection \ref{subsec:soliton_gas}, and is given by 
\begin{equation}
    v_a^\text{eff}\left[\{\rho_b\}\right](\theta)=\frac{\left(\partial_\theta e_a\right)^\text{dr}(\theta)}{\left(\partial_\theta p_a\right)^\text{dr}(\theta)}\,.
\end{equation}
Here, the superscript `dr' denotes the so-called dressing operation which is defined for any one-particle quantity $\omega_a(\theta)$ as
\begin{equation}
\label{eq:dressing}
    \omega_a^{\text{dr}} = \eta_a \left( \omega_a + \sum_b \Phi_{ab} * \vartheta_b \omega_b^{\text{dr}} \right)\,,
\end{equation}
where
 \begin{equation}
     \vartheta_a(\theta)=\frac{1}{1+e^\epsilon_a(\theta)}=\frac{\rho_a}{\rho^\text{tot}_a}
 \end{equation}
is the filling fraction. 

Exploiting the completeness of the charges, one arrives at the GHD equation \cite{2016PhRvX...6d1065C,2016PhRvL.117t7201B} 
\begin{equation}
    \partial_t\rho_a(x, t,\theta)+\partial_x \left(v_a^\text{eff}\left[\{\rho_b\}\right](\theta) \,\rho_a(x, t,\theta)\right)=0
\label{eq:general_GHD}
\end{equation}
for the  densities $\rho_a(t, x,\theta)$ of quasi-particle species $a$ and rapidity $\theta$ that are space and time-dependent on the Euler scale. 

Note that $v_a^\text{eff}$ carries an implicit dependence on $t$ and $x$ via the densities $\{\rho_j\}$ used to dress the derivatives of energy and momentum. These equations are supplemented by the dressing equations \eqref{eq:dressing} and the density equations \eqref{eq:BA_with_densities} necessary to reconstruct the filling fractions needed for the dressing from the quasi-particle (root) densities $\rho_a$. 

\section{Diffusion in GHD}\label{sG_diffusion}

The number of collisions experienced by a quasi-particle undergoes fluctuations in the local equilibrium state, resulting in diffusive corrections \cite{2018PhRvL.121p0603D,2019ScPP....6...49D} which can be interpreted in terms of a Gaussian broadening of the quasi-particle trajectories \cite{2018PhRvB..98v0303G}. Diffusion leads to higher-derivative corrections to the hydrodynamic equations, representing subleading corrections to the leading ballistic behavior at the Euler scale. As a result, the expression \eqref{eq:jh} can be viewed as the first term in a derivative expansion of the currents.  Diffusive corrections are accounted for by terms depending on the spatial variation of the charges:
%\begin{multline}
\begin{align}
    \mathtt{j}_{h} =  &\sum_a \int \mathrm{d}\theta  h_a(\theta) \Bigg[\rho_a(\theta) v_a^{\textrm{eff}}(\theta)
    \nonumber\\
    &-\frac12 \int\mathrm{d}\alpha \,\sum_b \mathfrak{D}_{ab}[\{\rho\}](\theta,\alpha)\,\partial_x\rho_b(\alpha)
    \Bigg]\,,
\end{align}
leading to the diffusive GHD (Navier--Stokes) equation
\begin{equation}
     \partial_t\rho_a+\partial_x \left(v_a^\text{eff}\left[\{\rho\}\right] \,\rho_a\right)=\frac12 \sum_b\partial_x\left(\mathfrak{D}_{ab}\left[\{\rho\}\right]\cdot\partial_x\rho_b\right)\,,
\end{equation}
where we suppressed the arguments $(x,t,\theta)$ and the dot on the right-hand side denotes action with the integral kernel. 

\section{Effective low-temperature equations}

Here, we give explicit formulae for the low-temperature asymptotics of the solutions for the Bethe Ansatz densities, effective velocities and dressed kernels which are used in our calculations.

\subsection{Reflectionless points}

At the reflectionless points $\xi=\frac{1}{N_B-1}$, the only relevant scattering events for the charge and momentum Onsager matrices involve the kinks $S$ (antikinks have equal contribution) and first breather type $B_1$. 
The relevant thermodynamic solitonic quantities read
\begin{align}
    \rho_S (\theta) &= \frac{m_S \cosh \left( \theta \right)}{2 \pi} \exp \left( -\frac{m_S}{T} \cosh\left( \theta\right)\right) \nonumber\\
    \rho_{S}^{\mathrm{tot}} (\theta) &= \frac{m_S \cosh \left( \theta \right)}{2 \pi} \nonumber\\
    v_{S}^{\mathrm{eff}} (\theta) &= \tanh \left( \theta \right) \nonumber\\
    q_{S}^{\mathrm{dr}} (\theta) &= 1 \nonumber\\
    p_{S}^{\mathrm{dr}} (\theta) &= m_S \sinh \left( \theta \right)
\end{align}
while for the first breather
\begin{align}
    \rho_{B_1} (\theta) &= \frac{m_{B_1} \cosh \left( \theta \right)}{2 \pi} \exp\left(-\frac{m_{B_1}}{T} \cosh\left( \theta\right) \right) \nonumber\\
    \rho_{B_1}^{\mathrm{tot}} (\theta) &= \frac{m_{B_1} \cosh \left( \theta \right)}{2 \pi} \nonumber\\
    v_{B_1}^{\mathrm{eff}} (\theta) &= \tanh \left( \theta \right) \nonumber\\
    q_{B_1}^{\mathrm{dr}} (\theta) &= 0\nonumber\\
    p_{B_1}^{\mathrm{dr}} (\theta) &= m_{B_1} \sinh \left( \theta \right)
\end{align}
where the first breather mass is
$m_{B_1} = 2 m_S  \sin \left( \pi \xi /2 \right)$.
The relevant dressed scattering kernels take the form
\begin{align}
    \Phi_{B_{1} B_{1}}^{\mathrm{dr}} (\theta) &= \frac{4 \sin\left( \pi \xi \right) \cosh\left( \theta \right)}{ \cos\left( 2 \pi \xi \right) - \cosh\left( 2 \theta \right)} \nonumber\\
    \Phi_{B_{1} S}^{\mathrm{dr}} (\theta) &= - \frac{4 \sin\left( \pi \xi /2 \right) \cosh\left( \theta \right)}{ \cos\left( \pi \xi \right) + \cosh\left( 2 \theta \right)}
\end{align}

\subsection{Attractive 2-magnon points}

At the couplings $\xi=\frac{1}{N_B + 1/2}$ where $N_B$ is the number of the breather species, the only relevant scattering events for the charge and momentum Onsager matrices involve the soliton $S$, the first breather $B_1$, and the two magnon species, the latter having equal contributions. The relevant thermodynamic solitonic quantities read
\begin{align}
    \rho_S (\theta) &= \frac{m_S \cosh \left( \theta \right)}{2 \pi} \exp \left( -\frac{m_S}{T} \cosh\left( \theta\right)\right) \nonumber\\
    \rho_{S}^{\mathrm{tot}} (\theta) &= \frac{m_S \cosh \left( \theta \right)}{2 \pi} \nonumber\\
    v_{S}^{\mathrm{eff}} (\theta) &= \tanh \left( \theta \right) \nonumber\\
    q_{S}^{\mathrm{dr}} (\theta) &= 0 \nonumber\\
    p_{S}^{\mathrm{dr}} (\theta) &= m_S \sinh \left( \theta \right)
\end{align}
while for the first breather
\begin{align}
    \rho_{B_1} (\theta) &= \frac{m_{B_1} \cosh \left( \theta \right)}{2 \pi} \exp\left(-\frac{m_{B_1}}{T} \cosh\left( \theta\right) \right) \nonumber\\
    \rho_{B_1}^{\mathrm{tot}} (\theta) &= \frac{m_{B_1} \cosh \left( \theta \right)}{2 \pi} \nonumber\\
    v_{B_1}^{\mathrm{eff}} (\theta) &= \tanh \left( \theta \right) \nonumber\\
    q_{B_1}^{\mathrm{dr}} (\theta) &= 0\nonumber\\
    p_{B_1}^{\mathrm{dr}} (\theta) &= m_{B_1} \sinh \left( \theta \right)
\end{align}
The magnonic quantities read
\begin{align}
    \rho_{M_2} (\theta) &= \frac{2} {\xi \cosh \left( \frac{2}{\xi}  \theta \right)} * \frac{\vartheta_S \left( \theta \right) m_S \cosh \left( \theta \right) \theta }{4 \pi ^2}  \nonumber\\
    \rho_{M_2}^{\mathrm{tot}} (\theta) &= \frac{\rho_{M_2} (\theta)}{2} \nonumber\\
    v_{M_2}^{\mathrm{eff}} (\theta) &= -\frac{T}{m_S \rho_{M_2}^{\mathrm{tot}} (\theta)} \partial_\theta  \left( \rho_{M_2} (\theta) \right)\nonumber\\
    q_{M_2}^{\mathrm{dr}} (\theta) &= 2\nonumber\\
    p_{M_2}^{\mathrm{dr}} (\theta) &= 0
\end{align}
The relevant dressed scattering kernels take the form
\begin{align}
    \Phi_{B_{1} B_{1}}^{\mathrm{dr}} (\theta) &= \frac{4 \sin\left( \pi \xi \right) \cosh\left( \theta \right)}{ \cos\left( 2 \pi \xi \right) - \cosh\left( 2 \theta \right)} \nonumber\\
    \Phi_{M_{2} S}^{\mathrm{dr}} (\theta) &= \frac{1}{2 \xi \cosh \left( \frac{2}{\xi} \theta \right)}
\end{align}

\end{document}